\documentclass[12pt]{article}
\usepackage{a4wide,graphicx,amsmath,amssymb,hyperref}
\usepackage[numbers,sort&compress]{natbib}
\usepackage{hypernat}
\usepackage{subfigure}
\usepackage{multirow}
\usepackage{axodraw}
\usepackage{color}
\usepackage{ulem}
\def\beq{\begin{equation}}
\def\eeq{\end{equation}}
\def\bea{\begin{eqnarray}}
\def\eea{\end{eqnarray}}

\newcommand{\newc}{\newcommand}

\newc{\stxt}[1]{{\color{magenta} #1}}

\newc{\scale}{0.85}

\newc{\half}{\frac{1}{2}}
\newc{\quart}{\frac{1}{4}}
\newc{\ifb}{\textrm{fb}^{-1}}
\newc{\fb}{\textrm{fb}}
\newc{\mass}{m}
\newc{\ord}{{\mathcal O}}
\newc{\Lag}{{\mathcal L}}
\newc{\sig}{\sigma}
\newc{\ct}{{\rm cos}\,\theta^*}
\newc{\ctsqr}{{\rm cos}^2\,\theta^*}
\newc{\ctquad}{{\rm cos}^4\,\theta^*}
\newc{\shat}{\hat{s}}
\newc{\MET}{{\not\!\!E_T}}

\renewcommand{\vec}[1]{\mathbf{#1}}
\newc{\vcross}{\times}
\newc{\vdot}{\cdot}

\newc{\Wp}{W^+}
\newc{\Wm}{W^-}
\newc{\Wpm}{W^{\pm}}
\newc{\gam}{\gamma}
\newc{\Wa}{W^+_1}
\newc{\Wb}{W^-_2}
\newc{\sla}{{\tilde{l}^+_1}}
\newc{\slb}{{\tilde{l}^-_2}}
\newc{\chia}{{\tilde{\chi}_1}}
\newc{\chib}{{\tilde{\chi}_2}}
\newc{\la}{l^+_1}
\newc{\lb}{l^-_2}
\newc{\nua}{\nu_1}
\newc{\nub}{\bar{\nu}_2}

\newc{\Vp}{V^+}
\newc{\Vm}{V^-}
\newc{\Fp}{F^+}
\newc{\Fm}{F^-}
\newc{\Sp}{S^+}
\newc{\Sm}{S^-}

\newc{\mw}{\mass_W}
\newc{\mnu}{\mass_{\nu}}
\newc{\mchi}{\mass_{\tilde{\chi}}}
\newc{\mchia}{\mass_{\chia}}
\newc{\mchib}{\mass_{\chib}}
\newc{\msl}{\mass_{\tilde{l}}}
\newc{\msla}{\mass_{\tilde{l}_1}}
\newc{\mslb}{\mass_{\tilde{l}_2}}
\newc{\mgg}{\mass_{\gamma\gamma}}

\newc{\maxmsl}{\msl^{\rm max}}
\newc{\maxmchi}{\mchi^{\rm max}}
\newc{\truemsl}{\msl^{\rm true}}
\newc{\truemchi}{\mchi^{\rm true}}

\newc{\dm}{\Delta m^2_{\tilde{l}\tilde{\chi}}}

\newc{\epsa}{\epsilon_{\gamma_1}}
\newc{\epsb}{\epsilon_{\gamma_2}}

\newc{\pw}{p_W}
\newc{\vpw}{\vec{\pw}}
\newc{\pwa}{p_{W1}}
\newc{\vpwa}{\vec{\pwa}}
\newc{\pwb}{p_{W2}}
\newc{\vpwb}{\vec{\pwb}}

\newc{\pli}{p_{l_i}}
\newc{\pla}{p_{l_1}}
\newc{\vpla}{\vec{\pla}}
\newc{\epla}{E_{l_1}}
\newc{\plb}{p_{l_2}}
\newc{\vplb}{\vec{\plb}}
\newc{\eplb}{E_{l_2}}

\newc{\pchii}{p_{\tilde{\chi}_i}}
\newc{\pchia}{p_{\tilde{\chi}_1}}
\newc{\vpchia}{\vec{\pchia}}
\newc{\pchib}{p_{\tilde{\chi}_2}}
\newc{\vpchib}{\vec{\pchib}}
\newc{\epchib}{E_{\tilde{\chi}_2}}

\newc{\psli}{p_{\tilde{l}_i}}
\newc{\psla}{p_{\tilde{l}_1}}
\newc{\vpsla}{\vec{\psla}}
\newc{\pslb}{p_{\tilde{l}_2}}
\newc{\vpslb}{\vec{\pslb}}

\newc{\pld}{P}

\newc{\pgg}{p_{\gamma\gamma}}
\newc{\vpgg}{\vec{\pgg}}
\newc{\epgg}{E_{\gamma\gamma}}

\newc{\pga}{p_{\gamma1}}
\newc{\vpga}{\vec{\pga}}
\newc{\pgb}{p_{\gamma2}}
\newc{\vpgb}{\vec{\pgb}}

\newc{\Lama}{\Lambda_{1}}
\newc{\Lamb}{\Lambda_{2}}
\newc{\Lamgg}{\Lambda_{\gamma\gamma}}
\newc{\Lamdm}{\Lambda_{\Delta}}
\newc{\Lammchi}{\Lambda_{\tilde{\chi}}}
\newc{\Lammsl}{\Lambda_{\tilde{l}}}
\newc{\Lamp}{\Lambda_{P}}

\newc{\vc}{\vec{C}}
\newc{\vd}{\vec{D}}
\newc{\ve}{\vec{E}}
\newc{\vf}{\vec{F}}

\newc{\eg}{e.g.~}
\newc{\ie}{i.e.~}
\newc{\cf}{cf.~}

\def\refeq#1{Eq.~(\ref{#1})}
\def\refsec#1{\ref{#1}}

\newc{\rp}{\mathcal{R}_p}
\newc{\ztwo}{\mathcal{Z}_2}
\newc{\emiss}{E_{\rm miss}}

\newc{\squark}{\tilde{q}}
\newc{\sneut}{\tilde{\chi}^0}
\newc{\slep}{\tilde{l}}

\newc{\phocep}{\texttt{PhoCEP}}

\begin{document}
  \titlepage
  \begin{flushright}
    Cavendish-HEP-2011-23
  \end{flushright}
  \vspace*{0.5cm}
  \begin{center}
    {\Large \bf Measuring the masses of a pair of semi-invisibly
      decaying particles in central exclusive production with forward
      proton tagging}\\
    \vspace*{1cm} \textsc{L.~A.~Harland-Lang,$^a$ C.~H.~Kom,$^{a,b,c}$
      K.~Sakurai$^{\,a,b,d,e}$ and W.~J.~Stirling$^{\,a}$}\\
    \vspace*{0.5cm} $^a$Cavendish Laboratory, University of Cambridge,
    Cambridge CB3 0HE, UK \\$^b$DAMTP, University of Cambridge,
    Cambridge CB3 0WA, UK\\$^c$Department of Mathematical Sciences,
    University of Liverpool, Liverpool L69 3BX, UK \\ $^d$Department
    of Physics, Nagoya University, Nagoya 464-8602, Japan
    \\$^e$Deutsches Elektronen-Synchrotron DESY, 22603 Hamburg,
    Germany
  \end{center}
  \vspace*{0.5cm}
  \begin{abstract}
    We discuss how the mass of new physics particles involved in a
    pair of short decay chains leading to two invisible particles, for
    example slepton pair production, followed by the decay into two
    leptons and two neutralinos, may be measured in central exclusive
    production (CEP) with forward proton tagging.  We show how the
    existing mass measurement strategies in CEP may be improved by
    making full use of the mass--shell constraints, and demonstrate
    that, with around 30 signal events, the masses of the slepton and
    neutralino can be measured with an accuracy of a few GeV.
  \end{abstract}

\section{Introduction}\label{sec:intro}

Many theories beyond the Standard Model (SM) include a dark matter
(DM) candidate as part of the mass spectrum.  Typically, the stability
of the DM candidate is the result of an assumed global $\ztwo$ parity,
under which the new (SM) particles are odd (even).  For example, in
the minimal supersymmetric extension of the SM (MSSM), this is the
so--called R--parity $\rp$, while in models with large extra
dimensions, this could be a Kaluza--Klein (KK) parity.

If such a DM candidate is produced at the Large Hadron Collider (LHC),
its presence could be revealed through the observation of large missing
transverse energy.  The $\ztwo$ parity implies that the DM candidate
will be pair--produced at the LHC.  This leads to complications in
measuring its properties, in particular its mass, since only the
vector sum of the transverse momenta of the two DM candidates in an
event can be measured.

Strategies to circumvent this problem abound in the literature~\cite{
  Paige:1996nx, Hinchliffe:1996iu, Allanach:2000kt, Lester:1999tx,
  Barr:2003rg, Cheng:2008hk, Barr:2007hy, Cho:2007dh, Nojiri:2008hy,
  Nojiri:2008vq, Burns:2008va, Nojiri:2010mk, Konar:2009wn,
  Barr:2009jv, Konar:2009qr, Ross:2007rm, Barr:2008ba, Barr:2011xt,
  Kawagoe:2004rz, Cheng:2007xv, Cheng:2008mg, Webber:2009vm,
  Nojiri:2010dk, Cheng:2011ya}.\footnote{ See also
  Ref.~\cite{Barr:2010zj} for a review. } Many of these assume the
existence of certain long decay chains.  An important example in the
MSSM is the pair production of squarks ($\squark$), followed by the
decay chain
\begin{eqnarray}\label{eq:standardchain}
  \squark &\to& q \sneut_2 \,,\nonumber \\ 
  &&\phantom{q}\sneut_2 \to l^{\pm} \slep^{\mp}\,, \nonumber \\
  &&\phantom{q\sneut_2 \to l^{\pm}} \slep^{\mp} \to l^{\mp} \sneut_1 \,,
\end{eqnarray}
where $\sneut_{1,2}$ are the neutralinos ($\sneut_1$ is the DM
candidate), and $\slep$ is a slepton.  The existence of more
constraints, \eg mass--shell conditions and dilepton invariant mass
edge measurements, than undetermined 4--momenta\footnote{In the
  present context, there are 6 ($=8-2$) unknowns: the four momenta of
  the two $\sneut_1$ (8), minus the measured missing transverse
  momenta (2).}  would allow simultaneous determination of the mass of
the new particles.  These long decay chains imply various challenges
for the relevant mass measurement techniques, \eg correct
identification of the decay topology and combinatoric problems in
assigning the identified final state particles to appropriate stages
of the decay chains, and so it is advantageous to consider alternative
methods to complement existing strategies.

The central exclusive production (CEP) of a system $X$ can be written
in the form
\begin{eqnarray}
 pp &\to& p\,+\,X\,+p\,,
\end{eqnarray}
where the `$+$' signs represent the presence of large rapidity
gaps. In this type of reaction, only the system $X$ is produced in the
central detector, with no additional hadronic activity in the absence
of pile--up, see for example Ref.~\cite{Khoze:2001xm,Albrow:2010yb}
(and references therein) for a review.  The outgoing protons scatter
at small angles and remain intact, so that additional kinematic
information can be obtained by installing proton tagging detectors far
from the interaction point~\cite{Albrow:2008pn}.  In particular, in
the presence of these detectors the momenta of the outgoing protons
can be measured precisely, and therefore the longitudinal momentum and
invariant mass of the central system X can be accurately determined.

In this paper, we present a mass measurement method utilising the
additional longitudinal kinematic information in the exclusive
two--photon production of a pair of charged particles ($X_1^+,X_2^-$),
\begin{eqnarray}
pp &\to& p\, +\, \gam\gam\,+\,p\,, \\ \nonumber
&&  \phantom{p\, +\,\,\,\,}  \gam\gam\to X_1^+X_2^-\,,
\end{eqnarray}
followed by the decay
\begin{eqnarray}
 X^{\pm} &\to& x^{\pm}+\chi\,,
\end{eqnarray}
where $x$ is a visible particle (or more generally, fully
reconstructed system of particles), and $\chi$ is an invisible
particle that is not seen in the detector. For concreteness, we will
consider the specific case of the exclusive two--photon production of
a pair of sleptons ($\sla,\slb$), followed by the decay into leptons
($\la,\lb$) and the lightest neutralinos ($\chia,\chib$) (note the
different notations compared with \refeq{eq:standardchain}), \ie
\begin{eqnarray}\label{eq:cepsusy}
pp &\to& p\, +\, \gam\gam\,+\,p\,, \nonumber \\ && \phantom{p\,
  +\,\,\,\,} \gam\gam\to\sla\slb\,,\nonumber \\ && \phantom{ p\,
  +\,\,\, \gam\gam\to} \sla \slb \to \la \lb \chia \chib\,,
\end{eqnarray}
leading to a signature with two tagged protons, an opposite--sign
same--flavour lepton pair and missing 4--momentum with no additional
jet activity.  The irreducible SM background comes from the production
of $W$ pairs, \ie
\begin{eqnarray}\label{eq:cepsm}
pp &\to& p\, +\, \gam\gam\,+\,p\,, \nonumber \\
&&  \phantom{p\, +\,\,\,\,}  \gam\gam\to\Wa\Wb\,,\nonumber \\
&& \phantom{ p\, +\,\,\,  \gam\gam\to} \Wa \Wb \to \la \lb \nua  \nub\,.
\end{eqnarray}

Compared to the long decay chain discussed in
\refeq{eq:standardchain}, the two--photon production process benefits
from a clean, purely leptonic signature, with no combinatoric
ambiguity.  Furthermore, assuming the slepton only decays into the
lightest neutralino, the process is essentially a two scale problem
that depends on the (true) masses of the slepton and the lightest
neutralino, $(\truemsl, \truemchi)$, together with discrete parameters
such as the particles' charges and spins.

The irreducible SM background is also under good experimental control.
This is because in the case of the signal, two--photon production must
create two same--flavour sleptons.  In the absence of large lepton
flavour violation, this will lead to two same--flavour leptons in the
final state.  For the SM background there is no such constraint on the
flavour of the final--state leptons: a measurement of the opposite
flavour final states $e^{\pm}\mu^{\mp}$ would therefore allow a cross
check of the background determination.

However, the cross section for the exclusive two--photon production of
slepton pairs is small when compared to the inclusive production cross
section from hard inelastic scattering, which is roughly two orders of
magnitude larger~\cite{delAguila:1990yw,Baer:1993ew}: as a result,
mass measurement in this central exclusive channel is only realistic
at the 14 TeV LHC for $\msl$ up to about 200 GeV.  However if the
coloured sparticles, \ie the squarks and the gluinos, are too heavy to
be produced copiously at the LHC, the presence and properties of light
colour singlet sparticles may be best measured via two--photon
production.  If (as we hope) the coloured sparticles are produced in
abundance, and long decay chains are observed, the present method
might still be used as an independent check or to provide external
constraints on the long decay chains.

The system in Eq.~(\ref{eq:cepsusy}) has four unknowns: only the
vector sum of the 4--momenta of $\chia$ and $\chib$ is measurable.
Given a consistent mass hypothesis\footnote{The technical definition
  of a consistent mass hypothesis will be given in the next section.}
$(\msl,\mchi)$, the four quadratic mass shell constraints allow the
system of equations to be solved analytically, up to a two--fold
degeneracy.  As a result, a $(\msl,\mchi)$ mass region consistent with
the measured 4--momenta can be obtained on an event--by--event basis.
We believe this consistent mass regions contains all information that
can be derived from the measured 4--momenta.  Neglecting finite
detector resolution effects, the (different) consistent mass region
from each event must contain the true mass point
$(\truemsl,\truemchi)$.  Combining the consistent regions from a
number of events will develop a peaking structure around this point,
hence in principle allowing determination of the true masses.

In determining the true masses and in separating the signal from SM
background, we find the maximum values of $\msl$ ($\maxmsl$) and
$\mchi$ ($\maxmchi$) within the consistent mass region particularly
useful.  This is because, by construction, $\maxmsl$ and $\maxmchi$
are bounded from below by $\truemsl$ and $\truemchi$ respectively.
They are also bounded from above by half the $\gam\gam$ invariant mass
$(\mgg)$ and half the missing energy $(\emiss)$.  An additional
advantage is the relative ease in constructing likelihood functions
from $\maxmsl$ and $\maxmchi$, which could be particularly useful
given a small data sample, as is likely the case in light of the small
two--photon production cross section.

The main aim of this paper is to study how well the slepton and
neutralino masses may be determined using $\maxmsl$ and $\maxmchi$
computed in each event.  These masses can also be estimated from
$\mgg$ and $\emiss$ \cite{Schul:2008sr,deFavereaudeJeneret:2009db},
which have threshold values at $2\truemsl$ and $2\truemchi$
respectively.  We shall compare the two methods and illustrate the
improvements that can be made by making use of the mass--shell
constraints.  The potential for measuring properties of the SM Higgs
boson~\cite{Khoze:1997dr,Khoze:2000cy}, long--lived
gluinos~\cite{Bussey:2006vx}, stops~\cite{Boonekamp:2005up}, MSSM
Higgs~\cite{Boonekamp:2005up,Heinemeyer:2007tu}, NMSSM
Higgs~\cite{Forshaw:2007ra} and other BSM Higgs
scenarios~\cite{Heinemeyer:2011jb, Belotsky:2004ex} in the central
exclusive mode with tagged forward protons have also been discussed in
the literature.

\begin{figure*}[!t]
  \begin{center}
    \scalebox{\scale}{
      \includegraphics{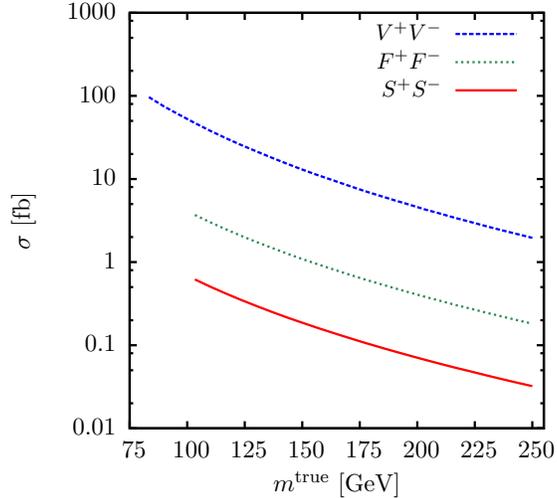}
    }
    \caption{Cross sections for pair production of spin--1
        ($V^+V^-$), spin--1/2 ($F^+F^-$) and spin--0 ($S^+S^-$)
        charged particles in central exclusive two--photon production
        as a function of the particle mass ($m^{\rm true}$) at the 14
        TeV LHC, calculated using the \texttt{PhoCEP} Monte Carlo event generator.}\label{fig:xsec}
  \end{center}
\end{figure*}

Note that the decay chain considered can be applied to new physics
models with different particle spins, for example the pair production
of charged spin--1/2 fermions followed by the decay into neutral
spin--1 bosons, as might be expected from a KK model.  In fact, it can
be shown that, given the same masses and charges, the production cross
section for spin--0 bosons is the smallest when compared with
spin--1/2 fermions and spin--1 bosons, see Fig.~\ref{fig:xsec}.  In
this paper, we shall concentrate on the first case, which is the most
conservative, and use the SUSY decay chain in \refeq{eq:cepsusy} as
our working example.

The paper is organised as follows.  In Section \refsec{sec:method}, we
present the analytic solutions consistent with the 4--momenta
measurable in \refeq{eq:cepsusy}, and derive the quantities $\maxmsl$
and $\maxmchi$.  We then present the numerical results in Section
\refsec{sec:numerical}, and compare between the performance of
$(\maxmsl,\maxmchi)$ and $(\mgg,\emiss)$ in determining
$(\truemsl,\truemchi)$, after including the CEP dynamics simulated by
a new event generator \phocep \cite{hepforge:phocep}.  This is
followed by an estimate of $(\truemsl,\truemchi)$ using the likelihood
method advertised above assuming an integrated luminosity of
$300\,\ifb$ at the 14 TeV LHC, which is the main result of our paper.
We conclude in Section \refsec{sec:summary}.

\section{The consistent solutions}\label{sec:method}

The 4--momenta of the particles in the decay chain in
\refeq{eq:cepsusy} are subjected to the constraints
\begin{eqnarray}\label{eq:constraints}
\pgg &=& \psla + \pslb \,,\nonumber \\
\psli &=& \pchii + \pli \,,\nonumber \\
\psli^2 &=& \msl^2 \,,\nonumber \\
\pchii^2 &=& \mchi^2\,,
\end{eqnarray}
where ($i=1,2$) and $\pgg$ is the 4--momentum vector sum of the two
$\gam$'s.  For simplicity, the sign of the sleptons and leptons are
dropped.  In the above expressions, only $\pgg$ and $\pli$ are
directly measurable.  The 4--momenta of the SUSY particles, $\psli$
and $\pchii$, can be expressed as linear combinations of $\pgg$,
$\pli$ and $\pld$, a {\it space--like} 4--momentum vector defined by
\begin{eqnarray}
\pld_{\mu} &\equiv& \epsilon_{\mu\nu\rho\sigma}\pgg^{\nu}\pla^{\rho}\plb^{\sigma}\,,
\end{eqnarray}
which has the property
\begin{eqnarray}
\pld\cdot\pgg\,=\,\pld\cdot\pla\,=\,\pld\cdot\plb\,=\,0\,.
\end{eqnarray}
\refeq{eq:constraints} implies
\begin{eqnarray}\label{eq:momentumconstraints}
\pchia &=& \half(1-a)\pgg-\half(b+1)\pla+\half(c-1)\plb -d\pld \,,\nonumber \\
\pchib &=& \half(1+a)\pgg+\half(b-1)\pla-\half(c+1)\plb +d\pld \,,\nonumber \\
\psla  &=& \half(1-a)\pgg-\half(b-1)\pla+\half(c-1)\plb -d\pld \,,\nonumber \\
\pslb  &=& \half(1+a)\pgg+\half(b-1)\pla-\half(c-1)\plb +d\pld \,,
\end{eqnarray}
with mass--shell conditions
\begin{eqnarray}\label{eq:massconstraints}
\mchia^2 &=& \quart(1-a)^2\mgg^2 + d^2\pld^2 \nonumber \\
&&-\half(1-a)(b+1)\pgg\cdot\pla + \half(1-a)(c-1)\pgg\cdot\plb-\half(b+1)(c-1)\pla\cdot\plb \,,\nonumber \\
\mchib^2 &=& \quart(1+a)^2\mgg^2 + d^2\pld^2 \nonumber \\
&&+\half(1+a)(b-1)\pgg\cdot\pla - \half(1+a)(c+1)\pgg\cdot\plb-\half(b-1)(c+1)\pla\cdot\plb \,,\nonumber \\
\msla^2 &=& \quart(1-a)^2\mgg^2 + d^2\pld^2 \nonumber \\
&&-\half(1-a)(b-1)\pgg\cdot\pla + \half(1-a)(c-1)\pgg\cdot\plb-\half(b-1)(c-1)\pla\cdot\plb \,,\nonumber \\
\mslb^2 &=& \quart(1+a)^2\mgg^2 + d^2\pld^2 \nonumber \\
&&+\half(1+a)(b-1)\pgg\cdot\pla - \half(1+a)(c-1)\pgg\cdot\plb-\half(b-1)(c-1)\pla\cdot\plb \,,
\end{eqnarray}
where the parameters $a$, $b$, $c$ and $d$ are to be determined. For
simplicity, we have neglected the lepton mass $m_l$ throughout,
although we note that consistent solution regions can also readily be
defined for general mass values.

Here $\mchi$ and $\msl$ need not be the true masses, but input
parameters consistent with a positive $d^2$.  The structure of the
system is such that upon subtracting the different expressions in
\refeq{eq:massconstraints}, the additive $d^2\pld^2$ term and the
terms bilinear or quadratic in $(a,b,c)$ drop out.  This leads to
three equations linear in $(a,b,c)$.  Defining the quantities
\begin{equation}
\Lamdm\equiv\frac{\dm}{\pla\cdot\plb}\,,\quad
\Lama\equiv\frac{\pgg\cdot\pla}{\pla\cdot\plb}\,,\quad
\Lamb\equiv\frac{\pgg\cdot\plb}{\pla\cdot\plb}\,,\quad
\Lamgg\equiv\frac{\mgg^2}{\pla\cdot\plb}\,,
\end{equation}
where
\begin{eqnarray}
\dm &\equiv& \msl^2-\mchi^2\,,
\end{eqnarray}
the solutions for the parameters $(a,b,c)$ are given by
\begin{eqnarray}\label{eq:solabc}
a&=&\Lamdm\frac{(\Lamb-\Lama)}{\Lamgg-2\Lama\Lamb} \,,\nonumber \\
b&=&1-\Lamb+\Lamdm\frac{\Lamgg-\Lamb(\Lama+\Lamb)}{\Lamgg-2\Lama\Lamb} \,,\nonumber \\
c&=&1-\Lama+\Lamdm\frac{\Lamgg-\Lama(\Lama+\Lamb)}{\Lamgg-2\Lama\Lamb} \,.
\end{eqnarray}
Inserting \refeq{eq:solabc} into \refeq{eq:massconstraints}, all four
mass constraints give
\begin{eqnarray}\label{eq:quad}
\Lammchi&=&c_a\Lamdm^2+c_b\Lamdm+c_c+d^2\Lamp \,,
\end{eqnarray}
where
\begin{equation}
  \Lammchi\equiv\frac{\mchi^2}{\pla\cdot\plb}\,,\quad
  \Lammsl\equiv\frac{\msl^2}{\pla\cdot\plb}\,,\quad
  \Lamp\equiv\frac{P^2}{\pla\cdot\plb}\,,
\end{equation}
and
\begin{eqnarray}
c_a&=&\frac{1}{4}\frac{(\Lama+\Lamb)^2-2\Lamgg}{\Lamgg-2\Lama\Lamb} \,,\nonumber \\
c_b&=&\frac{1}{2}(\Lama+\Lamb-2) \,,\nonumber \\
c_c&=&\frac{1}{4}(\Lamgg-2\Lama\Lamb)\,.
\end{eqnarray}

Note that in \refeq{eq:quad}, $c_a$, $c_b$, $c_c$ and $\Lamp$ are
constructed from directly measureable 4--momenta.  A mass hypothesis
$(\msl,\mchi)$ is consistent with the measured 4--momenta if it
results in a positive $d^2$, in which case a two--fold degenerate
solution for ($\pchia,\pchib,\psla,\pslb$) corresponding to $\pm |d|$
is obtained.

The boundary of consistent solutions on the $(\msl,\mchi)$ plane is
given by \refeq{eq:quad} by setting $d=0$.  To see this, note that
since $\pld$ is space--like, $\Lamp<0$.  This means that for a given
$\Lamdm$, $\Lammchi$ decreases as $d^2$ increases.  An upper boundary
on the $(\dm,\mchi^2)$ plane therefore has $d^2=0$.  The consistent
region is bounded from above by this quadratic curve, and below by the
$x$--axis, \ie $\mchi^2=0$, which can be transformed into a boundary
in the $(\msl,\mchi)$ plane.  Within the boundary, $d^2>0$.  By
construction, the solutions are time--like.

More observations on the properties of the consistent solutions can be
made.  First, the sign of the energy component of the two solutions
(for $\pm |d|$) must be the same, since it is always possible to boost
to a frame where the energy component of the space--like vector $\pld$
is zero, in which case the two solutions have the same energies.
Second, since the consistent solutions are continuous functions of
$\mchi$ and $\msl$, it follows that the energies of all consistent
solutions must have the same sign.  The energies must then be
positive, due to the fact that the true masses must be a consistent
solution, and hence lies within the boundary.  Third, the solution is
clearly symmetric when exchanging $\Lama$ and $\Lamb$.  This means
that the boundary obtained for the mass--shell conditions for $\pchia$
and $\pchib$ must be the same, consistent with the fact that both
constraints lead to \refeq{eq:quad}.

As discussed in Section \ref{sec:intro}, the consistent mass region
may be characterised by the maximum values of $\mchi$ and $\msl$, \ie
$\maxmchi$ and $\maxmsl$.  They correspond to the
stationary values of $\Lammchi$ and $\Lammsl$ as functions of $\Lamdm$
in \refeq{eq:quad}.  The results are
\begin{eqnarray}\label{eq:maxmass}
  (\maxmchi)^2 &=& (\pla\cdot\plb)\times\left(c_c-\frac{c_b^2}{4c_a}\right)\,,\nonumber \\
  (\maxmsl)^2 &=& (\pla\cdot\plb)\times\left(c_c-\frac{(c_b+1)^2}{4c_a}\right)\,.
\end{eqnarray}
By construction, $\maxmchi$ and $\maxmsl$ are bounded from below by
the true $\mchi$ and $\msl$ respectively.

So far we have shown that analytic boundaries containing mass
hypotheses consistent with the measured 4--momenta can be obtained on
an event--by--event basis.  In the next section, we present examples
showing the densities of consistent solutions given a large number of
events, and show the extent to which the quantities $\maxmsl$ and
$\maxmchi$ can be used to determine the true masses, crucially in a
small event sample.

\section{Numerical study}\label{sec:numerical}

Our numerical study is performed using the new Monte Carlo (MC) event
generator \phocep \cite{hepforge:phocep}. It simulates the two--photon
exclusive production of a $W^+W^-$ pair and a charged scalar pair,
$S^+S^-$, via the decay chains given in Eqs.~(\ref{eq:cepsusy}) and
(\ref{eq:cepsm}), as well as the production of two charged fermions,
$F^+F^-$, although we do not discuss this process in detail in this
paper.  The equivalent photon
approximation~\cite{Budnev:1974de,Albrow:2008pn} is used to calculate
the photon luminosity ${\rm d}L_{\gamma\gamma}/{\rm d} W$ due to
photon emission from the incoming protons, and this is then combined
with the known $\gamma\gamma\to X$ ($X=W^+W^-$, $S^+S^-$, $F^+F^-$)
subprocess cross sections to give the full CEP cross section
\begin{eqnarray}
 \frac{{\rm d}\sigma_{pp\to pXp}}{{\rm d}\Omega}&=&\int\frac{{\rm
     d}\sigma_{\gamma\gamma\to X}(W)}{{\rm d}\Omega}\frac{{\rm
     d}L^{\gamma\gamma}}{{\rm d}W}{\rm d W}\;,
\end{eqnarray}
where $W$ is the $\gamma\gamma$ cms energy. The produced particles are
then decayed according to the decay chains given in
Eqs.~(\ref{eq:cepsusy}) and (\ref{eq:cepsm}), and in the case of
$W^+W^-$ production full spin correlations for the subsequent decay
are included. The MC generates unweighted events, and full kinematic
cuts on the final state particles and outgoing protons can readily be
implemented. The survival factor $S^2$, representing the probability
that the outgoing protons do not undergo soft rescattering and
therefore spoil the exclusivity of the final state, is estimated to be
roughly 90\% for two--photon exclusive production~\cite{Khoze:2001xm},
and for simplicity is not included in the MC. The differential cross
sections for the $\gamma\gamma \to X$ subprocesses are summarised in
Appendix~\refsec{sec:appendix}.

In the following, we present numerical results of the methods
discussed in Section \ref{sec:method}.  For concreteness, we focus on
a SUSY model
\begin{itemize}
  \item SUSY model: $(\truemsl,\truemchi)=(150,100)$ GeV
\end{itemize}
and compare with results for SM $\Wp\Wm$ pair production, which
constitutes an irreducible background to the SUSY process.  For the
SUSY model, the slepton is assumed to always decay into a lepton and a
neutralino.  The branching fraction of the SM $W$ into one lepton
flavour is taken as 10.8\%~\cite{Nakamura:2010zzi}.

\begin{table}[!t]
  \centering
  \begin{tabular}{|c|c|c|}
    \hline
    &Before cuts ($\fb$) & After cuts ($\fb$) \\
    \hline
    SM&1.00&0.72\\
    SUSY&0.15&0.13\\
    \hline
  \end{tabular}
  \caption{Cross sections for the SUSY model with
    $(\truemsl,\truemchi)=(150,100)$ GeV and the SM background at the
    14 TeV LHC before and after the lepton cuts $-2.5 < \eta_{l} <
    2.5$ and $p_{T} > 10$ GeV. Cuts on the fractional momentum loss of
    the outgoing protons $0.0015<\xi_i<0.15$, are applied in both
    cases.  Branching fraction of the $W$ into one lepton flavour is
    included.  }
  \label{tab:xsec}
\end{table}

In our simulation, a final state proton is assumed to be tagged with
100\% efficiency if the fractional momentum loss lies within the range
\begin{itemize}
  \item $0.0015<\xi_i<0.15$\; $i=1,2$\,,
\end{itemize}
which corresponds to the expected coverage under the assumption that
roman pot detectors are installed at a distance of 220 and 420m from
interaction point, see for example Ref.~\cite{Chapon:2009hh}.

The following lepton pseudo--rapidity ($\eta_{l}$) and transverse
momentum ($p_{T}$) cuts
\begin{itemize}
  \item $-2.5 < \eta_{l} < 2.5$\,,
  \item $p_{T} > 10$ GeV\,,
\end{itemize}
are applied to the opposite--sign same--flavour lepton pairs.  We
assume 100\% detection efficiency if a lepton passes the above cuts.
The cross sections before and after the lepton cuts at the 14 TeV LHC
are displayed in Table~\ref{tab:xsec}.

\begin{figure*}[!t]
  \begin{center}
    \scalebox{\scale}{
      \includegraphics{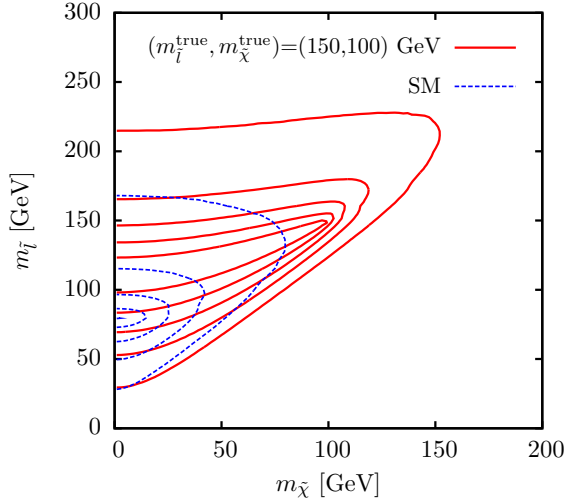}
    }
    \caption{Density of the consistent mass regions for a SUSY (solid
      red) and SM (dashed blue) sample.  From the outermost to the
      innermost, each set of contours delineate parameter regions with
      probability density larger than 0.1, 0.3, 0.5, 0.7 and 0.9.  The
      mass parameters for the SUSY model are
      ($\truemsl,\truemchi)=(150,100)$ GeV.  No detector smearing is
      included.}\label{fig:density}
  \end{center}
\end{figure*}

In Fig.~\ref{fig:density}, the density of the consistent
$(\msl,\mchi)$ regions are shown.  In both the SUSY and SM samples, a
peaking structure located in the vicinity of the true mass point
develops.  Finite detector resolution effects will smear the spike
structure slightly.  However, given the LHC's ability to measure
accurately the 4--momenta of the forward protons (in the presence of
proton taggers) and the charged leptons, the smearing effects should
be relatively mild.  The position of the high density regions are
significantly different such that the SUSY signal can be easily
separated from the background.

\begin{figure*}[!t]
  \begin{center}
    \scalebox{\scale}{
      \includegraphics{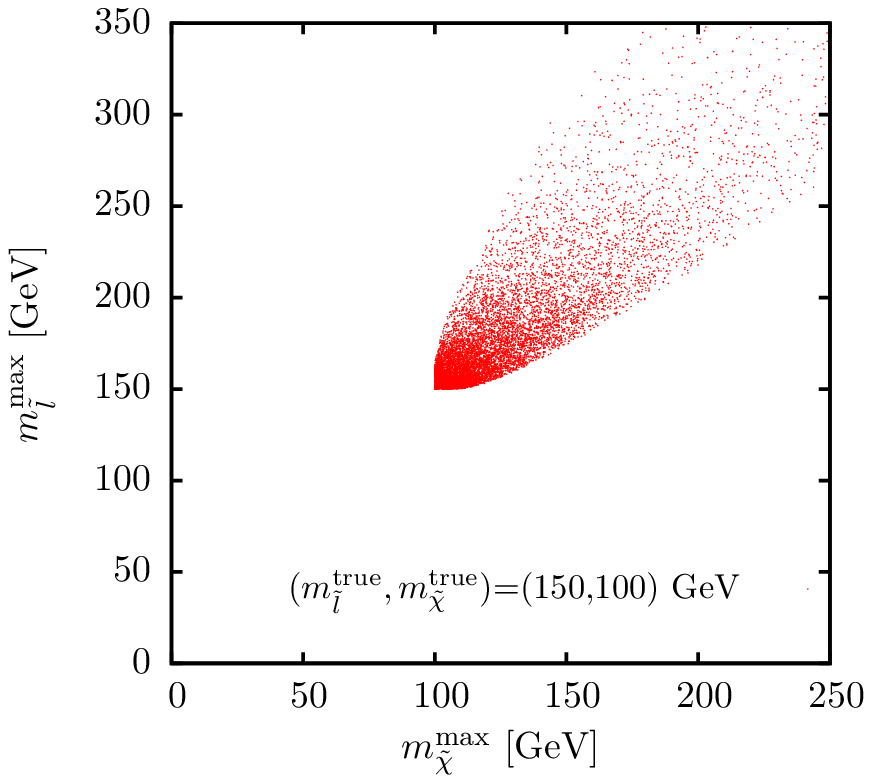}
      \includegraphics{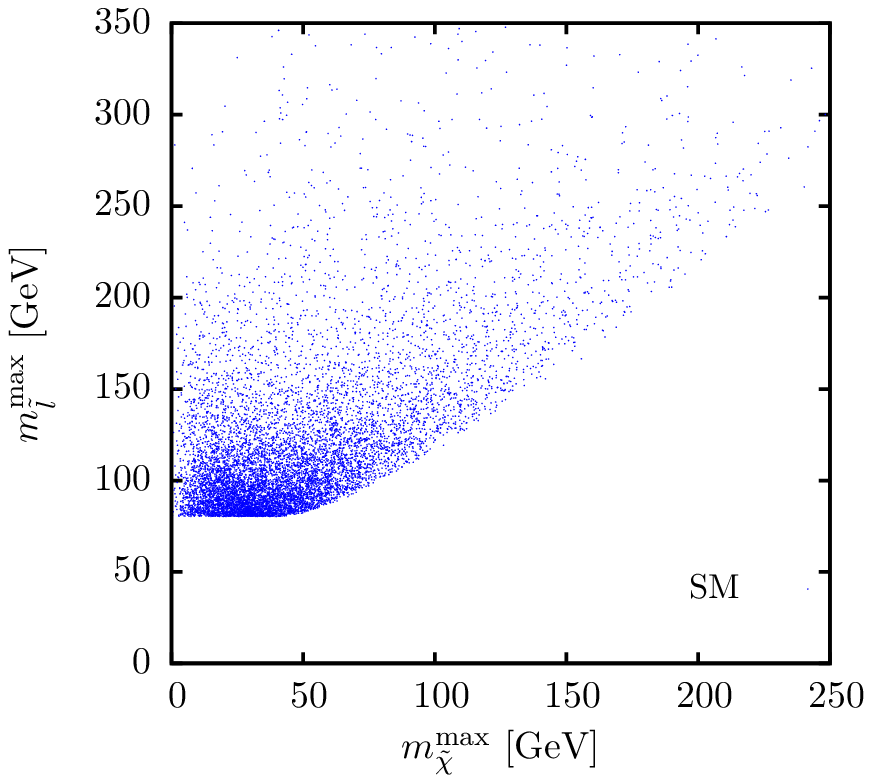}
    }
    \caption{Scatter plots showing the $(\maxmsl,\maxmchi)$ values
      consistent with each event.  Left: SUSY
      $(\truemsl,\truemchi)=(150, 100)$ GeV.  Right: SM.  Each
        plot contains 10,000 events.  No detector smearing is
      included.}\label{fig:scatter}
  \end{center}
\end{figure*}

Next we show in Fig.~\ref{fig:scatter}~scatter plots for $\maxmchi$
and $\maxmsl$, each of which contains 10,000 events.  Recall that, by
construction these two quantities are bounded from below by
$\truemchi$ and $\truemsl$.  As we can see, in both the SUSY and SM
samples $(\maxmsl,\maxmchi)$ cluster close to the true mass point
$(\truemsl,\truemchi)$.  This indicates that a good estimation of the
true mass parameters and identification of the presence of a SUSY
signal from the SM background could in principle be achievable.

To show the enhancement obtained by utilising the additional kinematic
information, in particular in mass determination, we compare our
results with an alternative method which measures the end points of
the missing energy ($\emiss$) and $\mgg$ distributions
\cite{Schul:2008sr}.  The end point values of these distributions are
at $2\truemchi$ and $2\truemsl$ respectively.  These distributions
take into account the sum of longitudinal momenta and energy of the
neutralinos that can be measured in CEP with tagged forward protons,
but do not include information from mass constraints.

\begin{figure*}[!t]
  \begin{center}
    \scalebox{\scale}{
      \includegraphics{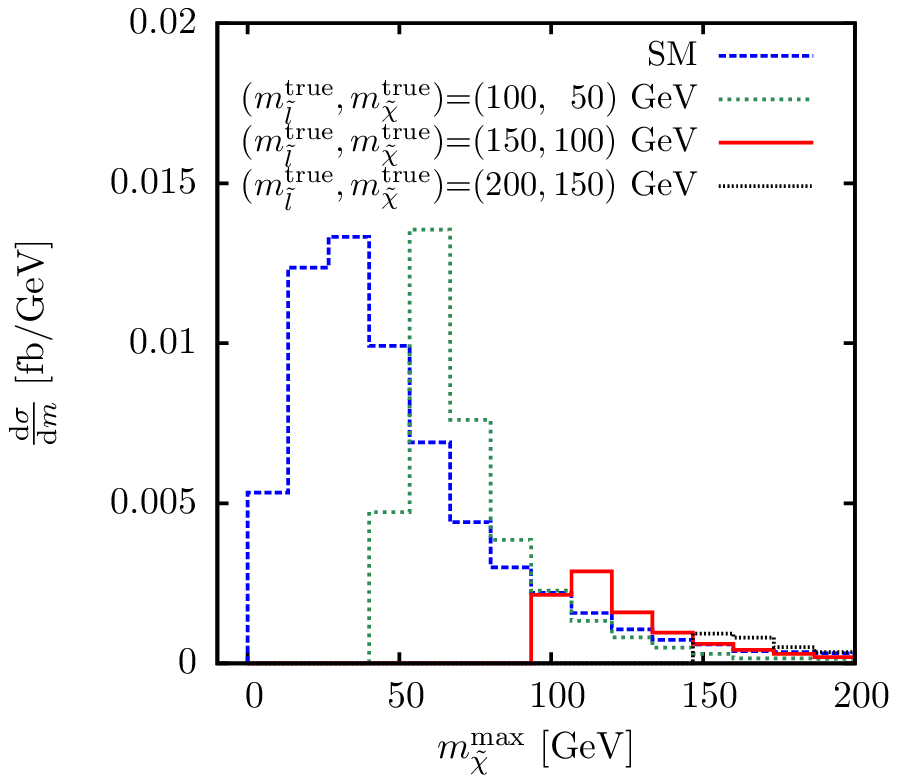}
      \includegraphics{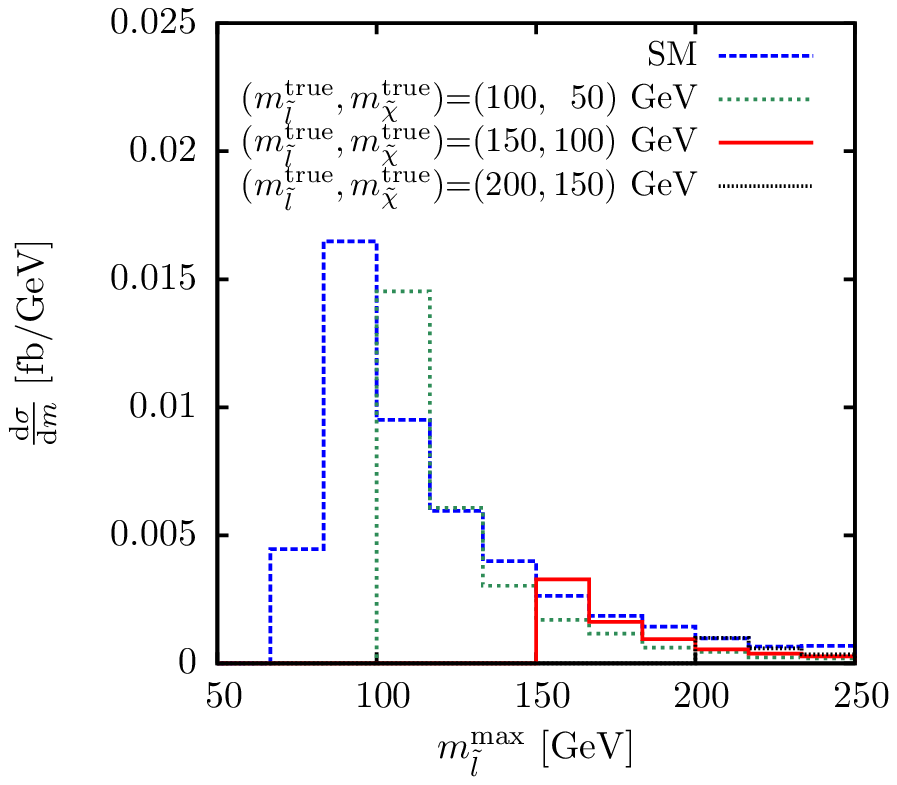}
    }
    \caption{Histograms showing the $\maxmchi$ (left) and $\maxmsl$
      (right).
      No detector smearing is included.}\label{fig:maxmass_h}
  \end{center}
\end{figure*}

\begin{figure*}[!t]
  \begin{center}
    \scalebox{\scale}{
      \includegraphics{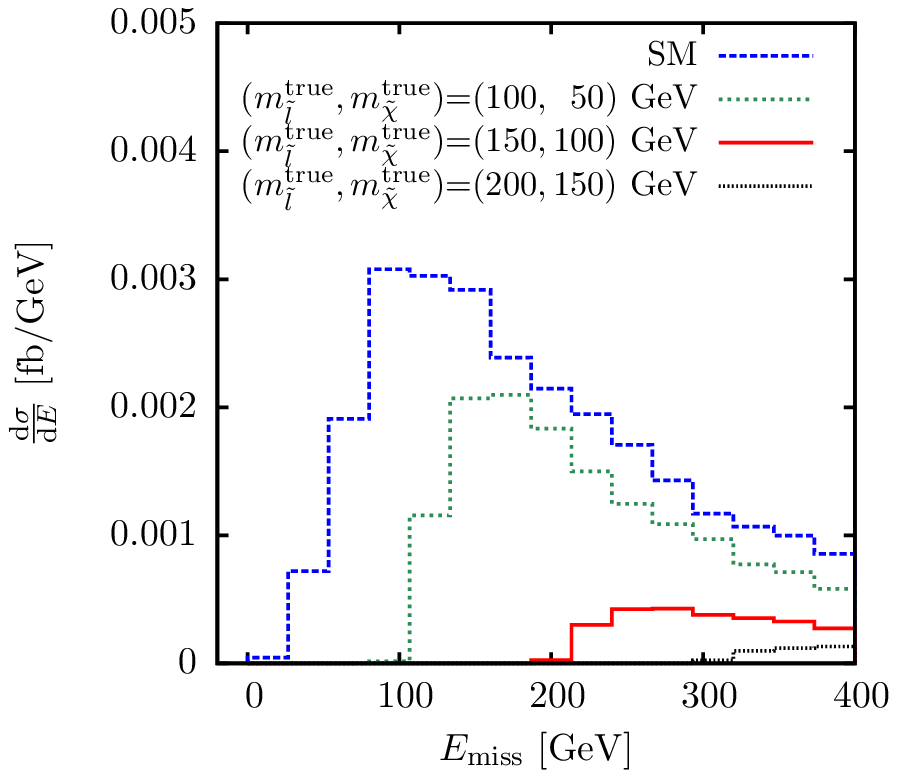}
      \includegraphics{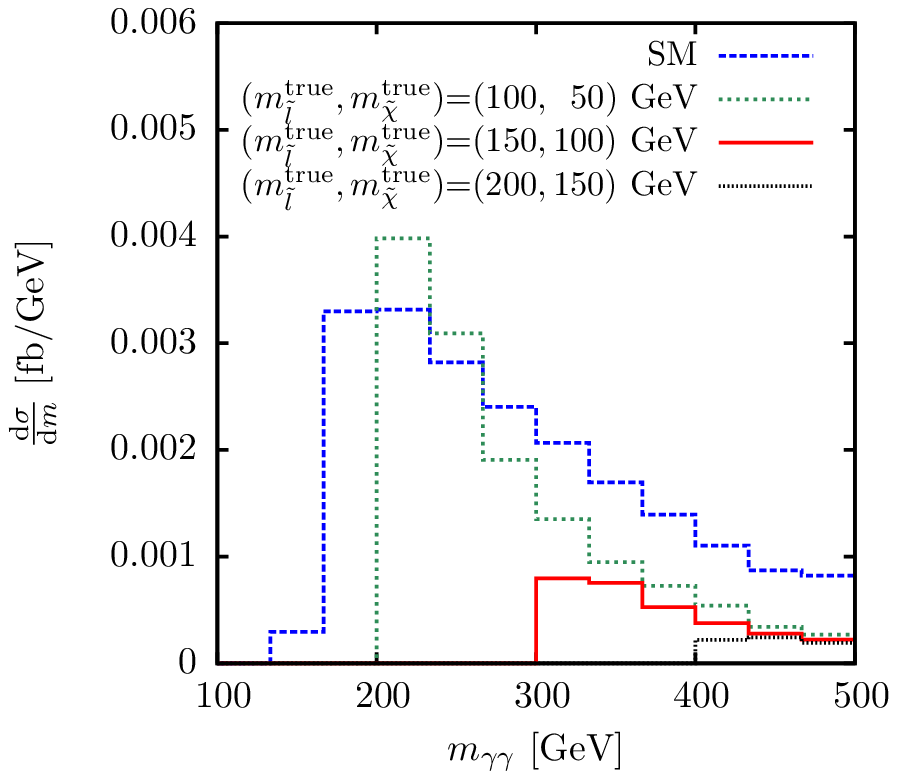}
    }
    \caption{Histograms showing the $E_{\rm miss}$ (left) and
      $m_{\gam\gam}$ (right) distributions.  No detector smearing
      included.}\label{fig:EmissMgg}
  \end{center}
\end{figure*}

In Fig.~\ref{fig:maxmass_h}, we show the $\maxmchi$ and $\maxmsl$
distributions.  The $\emiss$ and $\mgg$ distributions are displayed in
Fig.~\ref{fig:EmissMgg}.  Compared to the $\maxmchi$ and $\maxmsl$
distributions, the $\emiss$ and $\mgg$ distributions are generally
broader.  This qualitative difference is due to the fact that both
$\emiss/2$ and $\maxmchi$ ($\mgg/2$ and $\maxmsl$) are bounded from
below by $\truemchi$ ($\truemsl$), while $\maxmchi$ ($\maxmsl$) is in
addition bounded from above by $\emiss/2$ ($\mgg/2$).  To compare the
difference between these two methods over a range of masses, we show
two additional SUSY models, specified by
$(\truemsl,\truemchi)=(100,50)$ GeV and
$(\truemsl,\truemchi)=(200,100)$ GeV in addition to the `reference'
SUSY model and the SM background.  We see that, compared with the
$(\emiss,\mgg)$ distributions, the $(\maxmchi,\maxmsl)$ distributions
benefit from a better signal--to--background ratio (S/B) in the signal
region, which improves as the mass scale increases.  Here, the signal
region for a distribution is defined as the region above the lower
threshold value determined by $(\truemsl,\truemchi)$. For example, for
$(\truemsl,\truemchi)=(150,100)$ GeV, the S/B ratio in the signal
region is approximately 1 for both $\maxmsl$ and $\maxmchi$
distributions, while it is approximately $1/4$ and $1/3$ for the
$\emiss$ and $\mgg$ distributions respectively.  Importantly, the
distribution of the {\it signal} $\emiss$ distributions near the
endpoint becomes increasingly smooth as the mass scale increases,
leading to possibly large uncertainties in determining the endpoint
value.  In contrast, the impact on the $\maxmchi$ distribution is
relatively mild.

To estimate the precision in determining $(\truemsl,\truemchi)$ that
can be obtained by measuring $(\maxmsl,\maxmchi)$ at the LHC, we
perform a pseudo--experiment which includes detector smearing effects.
The 4--momenta of the final state leptons are smeared via
\begin{eqnarray}\label{eq:leptonsmearing}
\left(\frac{\delta E}{E}\right)^2 &=& \frac{a^2}{E} + b^2\,,
\end{eqnarray}
where $a=0.1\,\sqrt{\rm GeV}$ and $b=0.007$
\cite{CMSTDRII,ATLASsmearing}.  The fractional momentum loss $\xi$ of
the final state protons is smeared via\footnote{We choose this level
  of smearing to reproduce the quoted `missing mass' resolution of
  roughly $2-3$ GeV for a $120$ GeV SM Higgs
  boson~\cite{Albrow:2008pn}}
\begin{eqnarray}\label{eq:pxsmearing}
  \frac{\delta \xi}{\xi} &=& 0.04\,.
\end{eqnarray}

In the pseudo--experiment, the number of events we generate for both
the SUSY signal, $(\truemsl,\truemchi)=(150,100)$ GeV, and SM
background is based on the cross sections displayed in
Table~\ref{tab:xsec}, assuming an integrated luminosity of $300\,{\rm
  fb}^{-1}$, \ie 216 and 38 events for the SM and SUSY model
respectively.  We further define the signal region as the range within
$\maxmsl=[130:230]$ GeV and $\maxmchi=[80:180]$ GeV to increase the
S/B ratio.  In our particular pseudo--experiment sample, 24 and 36
events are observed in the signal region for the SM and SUSY model
respectively.

We construct the posterior probability density of ($\msl,\mchi$) and
identify the 68\% and 95\% credibility region as follows.  Given a
hypothesis ${\bf m} = (\msl,\mchi)$, the probability density function
(p.d.f.) for the SUSY sample in obtaining ${\bf m}^{\rm max} =
(\maxmsl,\maxmchi)$, denoted by $\rho_{\rm SUSY}({\bf m}^{\rm
  max};{\bf m})$, is estimated by generating 1 million events, and
binning the events passing the cuts into a grid with bin size $2\,{\rm
  GeV}\times 2\,{\rm GeV}$ on the $(\maxmsl,\maxmchi)$ plane.  We
generate 1189 SUSY mass points in the range of $\msl=[145:152]$ GeV
and $\mchi=[93:103]$ GeV with a 0.25\,GeV step size.  The p.d.f. for
the SM background sample, $ \rho_{\rm SM}({\bf m}^{\rm max})$, is
estimated by generating 1 million events in the same way.  The
combined probability density, $\rho({\bf m}^{max};{\bf m})$, is then
given by
\begin{eqnarray}\label{eq:density}
  \rho({\bf m}^{\rm max};{\bf m}) &=& f_{\rm SM}\rho_{\rm
    SM}({\bf m}^{\rm max}) + f_{\rm SUSY}\rho_{\rm
    SUSY}({\bf m}^{\rm max};{\bf m})\,,
\end{eqnarray}
where $f_{\rm SM}$ and $f_{\rm SUSY}$ are the fraction of SM and SUSY
cross section after the cuts,
with $f_{\rm SM}+f_{\rm SUSY}=1$.

The likelihood function, i.e. joint p.d.f., may be written as
\begin{eqnarray}\label{eq:likelihood}
  L({\bf m}) &=& \prod_{i}\rho({\bf m}^{\rm max};{\bf m})\,,
\end{eqnarray}
where $i$ runs over the number of events observed in the signal
region.  We define the posterior probability density as\footnote{ In
  other words, we take a constant prior p.d.f., $\pi({\bf m})=c$, in
  the Bayesian posterior p.d.f.
$$ p({\bf m}) = \frac{L({\bf m}) \pi({\bf m})}{\int d{\bf m}' L({\bf
      m}') \pi({\bf m}') }\,.$$.}
\begin{eqnarray}\label{eq:posterior}
  p({\bf m}) &=& \frac{L({\bf m})}{\int d{\bf m}' \, L({\bf m}')}\,.
\end{eqnarray}
A hypothesis ${\bf m}$ is inside the 68\% (95\%) credibility region if
\begin{eqnarray}\label{eq:cr_region}
\int d{\bf m'}\, p({\bf m'}) < 0.68~(0.95),
\end{eqnarray}
where the integration is carried out over ${\bf m'}$ satisfying
$p({\bf m'}) \ge p({\bf m})$.

\begin{figure*}[!t]
  \begin{center}
    \scalebox{1.1}{
      \includegraphics{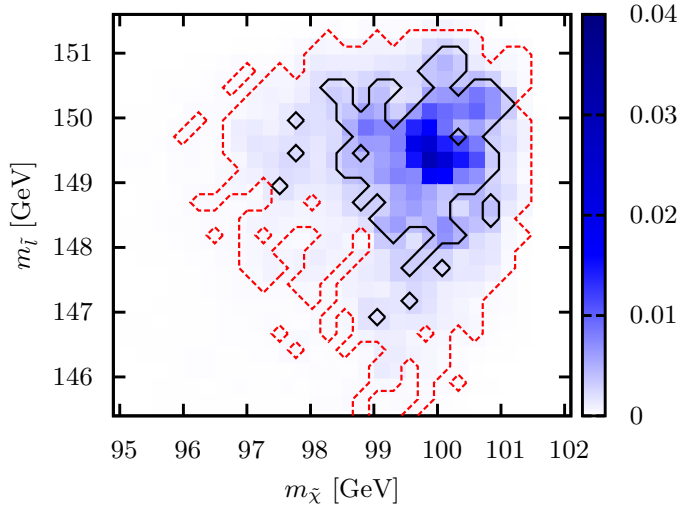}
    }
    \caption{Credibility distribution for the $(\msl,\mchi)$
      hypothesis in a pseudo--experiment.  The mass parameters of the
      SUSY model are given by $(\truemsl,\truemchi)=(150,100)$ GeV.
      The solid black and dashed red contours contain the 68\% and
      95\% credibility regions respectively.}\label{fig:likelihood}
  \end{center}
\end{figure*}

We show the $p({\bf m}) \Delta \msl \Delta \mchi$ distribution
($\Delta \msl = \Delta \mchi = 0.25$\,GeV) from an event sample from
one pseudo--experiment in Fig.~\ref{fig:likelihood}.  In this
particular sample, with 60 events (24 SM background, 36 SUSY signal)
passing the selection criteria, $(\truemsl,\truemchi)$ can be
estimated to about 3--4 GeV accuracy at 95\% credibility.  Results
from other event samples, which assume the same integrated luminosity,
show similar level of accuracies.  By applying the likelihood
calculation on event samples which do not include finite detector
resolution effects, we find that the uncertainty is primarily due to
statistical fluctuations.

As can be seen in Fig.~\ref{fig:likelihood}, there is a preference
towards the lower mass hypothesis.  This is a consequence of the
peaking structure of the $(\maxmsl,\maxmchi)$ distributions just above
$(\truemsl,\truemchi)$.  This implies that a hypothesis $(\msl,\mchi)$
where $\msl>\truemsl$ and $\mchi>\truemchi$ is likely to have many
signal events in the region where $\rho_{\rm SUSY}({\bf m}^{\rm
  max};{\bf m})$ is zero, and therefore they only `see' the likelihood
contribution from $f_{\rm SM}\rho_{\rm SM}(\maxmsl,\maxmchi)$, which
is rather low.  On the other hand the likelihood reduces less quickly
in the case where $\msl<\truemsl$ and $\mchi<\truemchi$ due to the
tail of the $\maxmsl$ and $\maxmchi$ distributions.

\section{Summary and discussion}\label{sec:summary}

In this paper, we have discussed how the additional kinematic
information obtained in CEP with tagged forward protons can be used to
simultaneously measure the masses of the particles involved in the
short decay chain displayed in Eq.~(\ref{eq:cepsusy}).  We presented a
full solution to the system, and found that the distributions of the
quantities $\maxmsl$ and $\maxmchi$, which are the maximum mass values
consistent with a given event, are particularly useful in estimating
their corresponding true mass values $\truemsl$ and $\truemchi$.  We
then performed a likelihood analysis to show that for our reference
SUSY model, $(\truemsl,\truemchi)=(150,100)$ GeV, the true masses can
be estimated to a few GeV at 95\% credibility with just
$\mathcal{O}(30)$ signal events.

As emphasised in Ref.~\cite{Bussey:2006vx}, the possibility to measure
central exclusive production in the high luminosity phase of the LHC
is crucial to a wide range of processes studied in the literature.
However, one potentially important issue which our simulation has not
included is the effect of pile--up: at high luminosity there are
expected to be multiple overlap events in each bunch crossing which
will fill the rapidity gaps between the central system and the
outgoing protons. Even so, by making use of the forward proton taggers
and the high precision time--of--flight detectors discussed in
Ref.~\cite{Albrow:2008pn} to demand that the tagged protons and the
opposite--sign lepton pair all point to the same origin, we may hope
to suppress the effect of pile--up, given good vertexing
resolution. However a more detailed investigation is beyond the scope
of the present study.

Nevertheless, we find the level of accuracies obtainable with the
small ($\ord(30)$) event sample encouraging.  The prospects for
measuring the masses in the corresponding decay chain in KK scenarios
are even better, due to the much higher signal cross section given the
same mass parameters, see Fig.~\ref{fig:xsec}.

As discussed in the Introduction, the production of a slepton pair,
followed by the decay into neutralinos and leptons in an inelastic
scattering is clearly possible.  As is well--known, the energy and the
longitudinal momentum of the hard process cannot be measured in a
hadron--hadron collision.  Compared to the CEP process that we
consider, the inelastic process has a much larger cross section.  In
principle, the MT2 method~\cite{Lester:1999tx} can be used to
determine the true masses by observing a kink structure at the true
mass value.  This kink structure is formed by events where the two
slepton system recoils against initial state radiation and has a
non--zero transverse momentum.  However the probability that the
system has a significant transverse momentum is very low and the kink
resides at the smooth tail of the distribution.  Once detector
smearing effects are included, it is likely that observing such a kink
structure will be difficult.  In this context, measuring the particle
masses in this particular decay chain in CEP with forward proton
tagging might be the best, if not the only, way.

Finally, we note that the mass measurement technique presented here
could also be used in future linear colliders.  A relevant process
could be
\begin{eqnarray}\label{eq:linear}
e^+e^- &\to& \sla \slb, \nonumber \\ && \sla \slb \to \la \lb \chia
\chib\,,
\end{eqnarray}
where the $\pgg$ is simply the vector sum of the incoming electron
4--momenta.  At parton level, the leptons have a flat energy
distribution with endpoints $E_{\rm max,min}$ at
\begin{eqnarray}
  E_{\rm max,min} &=&
  \frac{\sqrt{s}}{4}\left[1-\left(\frac{\truemchi}{\truemsl}\right)^2\right]\left[1\pm\sqrt{1-\left(\frac{2\truemsl}{\sqrt{s}}\right)^2}\right]\,,
\end{eqnarray}
\cf Ref.~\cite{Martin:1997ns}.  Since the masses are determined by the
endpoints, the events away from the endpoints are not directly used.
Our $\maxmsl$ and $\maxmchi$ distributions on the other hand peak
strongly near $\truemsl$ and $\truemchi$, so they are likely to
provide better statistics given the same number of events.  However we
postpone a comparative study of these two measurement methods to the
future.

\section*{Acknowledgements}

We thank members of the Cambridge SUSY Working Group, particularly
Chris Lester, for many useful discussions.  This work has been
supported in part by the Isaac Newton Trust and the STFC.  CHK would
like to thank the Particle and Astroparticle Physics group at MPIK
Heidelberg and the Rudolf Peierls Centre for Theoretical Physics at
Oxford for hospitality while part of the work was carried out.  KS is
supported in part by the YLC (Young Leaders Cultivation) programme in
Nagoya University.  LHL acknowledges financial support from the
University of Cambridge Domestic Research Studentship scheme. WJS and
LHL acknowledge financial support in the form of an IPPP
Associateship.

\appendix
\section{Differential QED cross sections with $\gam\gam$ initial states}\label{sec:appendix}
For ease of reference, the differential cross sections for pair
production of final state particles for $\Wp\Wm$\cite{Tupper:1980bw},
$\Sp\Sm$\cite{Ohnemus:1993qw} and $\Fp\Fm$\cite{Brodsky:1970vk} are
listed below:
\begin{eqnarray}
\frac{d\sig}{d\ct}(\gam\gam\to\Wp\Wm) &=& \frac{\pi\alpha^2\beta}{\shat}\frac{19-6\beta^2(1-\beta^2)+2(8-3\beta^2)\beta^2\ctsqr+3\beta^4\ctquad}{(1-\beta^2\ctsqr)^2} \nonumber \,, \\
\frac{d\sig}{d\ct}(\gam\gam\to\Sp\Sm) &=& \frac{\pi\alpha^2\beta}{\shat}\left[1-\frac{2(1-\beta^2)}{(1-\beta^2\ctsqr)}+\frac{2(1-\beta^2)^2}{(1-\beta^2\ctsqr)^2}\right] \nonumber \,, \\
\frac{d\sig}{d\ct}(\gam\gam\to\Fp\Fm) &=& \frac{2\pi\alpha^2\beta}{\shat}\frac{1+2\beta^2(1-\beta^2)(1-\ctsqr)-\beta^4\ctquad}{(1-\beta^2\ctsqr)^2}\,,
\end{eqnarray}
where $\beta=(1-4m^2/\shat)^{1/2}$, $(m=m_{W},m_{F},m_{S})$,
$\theta^*$ is the angle of out--going particles with respect to the
$\gamma$'s in the C.M. frame.  The charges and colour factors are all
assumed to be 1.  The total cross sections are given by
\begin{eqnarray}
\sig(\gam\gam\to\Wp\Wm) &=& \frac{\pi\alpha^2\beta}{\shat}\left[2\frac{22-9\beta^2+3\beta^4}{1-\beta^2}-3\frac{1-\beta^4}{\beta}\ln\left(\frac{1+\beta}{1-\beta}\right) \right] \nonumber \,, \\
\sig(\gam\gam\to\Sp\Sm) &=& \frac{2\pi\alpha^2\beta}{\shat}\left[2-\beta^2-\frac{1-\beta^4}{2\beta}\ln\left(\frac{1+\beta}{1-\beta}\right)\right] \nonumber\,, \\
\sig(\gam\gam\to\Fp\Fm) &=& \frac{4\pi\alpha^2\beta}{\shat}\left[-2+\beta^2+\frac{3-\beta^4}{2\beta}\ln\left(\frac{1+\beta}{1-\beta}\right) \right] \,.
\end{eqnarray}


\end{document}
